# A Geometrical Manifold of Entropy for Fisher Information and Quantum Potential

Germano Resconi
Catholic University, via Trieste 17 Brescia , Italy
resconi@numerica.it

Ignazio Licata
ISEM, Inst. For Scientific Methodology, PA, Italy
Ignazio.licata@ejtp.info

Davide Fiscaletti
SpaceLife Institute, San Lorenzo in Campo (PU), Italy
spacelife.institute@gmail.com

**Abstract**
It is here proposed a geometric approach for the problem of describing entropy in a quantum system. We make use of an extension of tensor calculus called morphogenetic calculus. By using such formalism we express the entropy of a quantum system as the superposition of Boltzmann entropies. In this way we also provide a reading of the relational interpretation of quantum mechanics. Moreover we show that the Bohm quantum potential emerges as a consequence of the classical equilibrium under the constraint of a minimum condition of Fisher information. In this way, Bohm quantum potential appears as a non-Euclidean deformation of the probabilistic space. Finally we investigate the possible quantum-relativistic extensions of the theory and the connections with the problem of quantum gravity.

**Introduction**
The advantage of a geometric representation of physical phenomena is not only the elegance, but also and above all an immediate visualization of processes. In this paper we extend the tensor calculus to operators represented by non-quadratic matrices and we propose a general form of geometrization called morphogenetic calculus. This mathematical formalism introduces interesting perspectives in the interpretation of some important results of quantum mechanics.

In the first part of the article, by making use of the morphogenetic calculus, we will consider the entropy of a quantum system as a vector of the superposition of many different entropies whose values are conditioned by the observer and will provide a new suggestive reading to Rovelli's relational quantum mechanics. Any observer makes the system in superposition collapse to a classical value of entropy, i.e. it fixes classical information by a measurement. When the observer does not interact with a quantum system, the composed system given by the quantum microsystem+observer (and thus, somewhat the universe) is in an entangled state. When the observer interacts with the microsystem under consideration thus becoming an active element, the composed system given by the quantum microsystem+observer is in a pure state and follows the classical thermodynamics defined by one specific entropy and, in correspondence, the measured variable of the system under consideration takes the measured value.

In the second part of the article, we change the reference of the entropic approach into a non Euclidean space where we obtain, by the covariant derivatives in a general form, the Bohm quantum potential by a minimum principle of the average action. In this approach, the Bohm quantum potential emerges as a consequence of the classical equilibrium condition in Boltzmann entropy, namely in the extreme condition of Fisher information and can be considered as an information medium determined by informational lines associated with the vector of the superpose entropies. In the entropic picture derived from the morphogenetic calculus, the vector of the superposition of different entropies indicates the informational lines of the quantum potential.

## 1. Tensor, covariant and contravariant derivatives and the morphogenetic system

In this chapter we introduce a simple extension of tensor calculus we will call morphogenetic calculus, a mathematical tool to individuate a geometric picture of the relations between Fisher information, Bohm quantum potential and Boltzmann entropy ( Resconi e Nikravesh, 2007).
Given the transformation.

$$y_j = y_j(x_1, x_2, ......, x_n),$$

$$j = 1, 2, ..., m,$$

$$n \geq m$$

we have

$$dy_k = J_{k,j} dx^j = \sum_j \frac{\partial y_k}{\partial x_j} dx_j \quad (1)$$

or

dy = J dx

*with*

$k = 1, 2, ..., m, j = 1, 2, ...., n$

$m \geq n$

In the morphogenetic theory $dx^j$ are the sources and $dy_k$ are the inputs. In the ordinary tensor calculus m = n and the sources are the contravariant components of the vector dy.
And the inputs $dy_k$ are the covariant components of dy. Now because J is a rectangular m x n matrix the computation of the sources can be obtained by the new equation

$$J^T dy = J^T J \, dx$$

$$dx = (J^T J)^{-1} J^T dy$$

In the ordinary tensor calculus J is a quadratic matrix so we have the tensor relation

$$dx^j = J^{-1} dy$$

$$dy_k = J^T dy$$

In the morphogenetic system we have

$$dx^j = J^+ dy = (J^T J)^{-1} J^T dy$$

where $J^+$ is the pseudo inverse of the matrix J

In the traditional tensor J is a quadratic matrix, so we have the expression

$$J dx^j = J J^+ dy = J(J^T J)^{-1} J^T dy = Q dy = dy$$

We can rebuilt the vector dy by the contravariant components. In the morphogenetic system Q is the projection operator on the space of the independent variables x.
Now the fundamental metric tensor is given by the expression

$$J^T J = \sum_i \frac{\partial y_i}{\partial x_h} \frac{\partial y^i}{\partial x_k} = a_{h,k}$$

And

$$(J^T J)^{-1}(J^T J) = \sum_{h,k} a^{h,k} a_{h,k} = \begin{bmatrix} 1 & 0 & \dots & 0 \\ 0 & 1 & \dots & 0 \\ \dots & \dots & \dots & \dots \\ 0 & 0 & \dots & 1 \end{bmatrix} = I.$$

So we have also the classical expression

$$V^k(x) = \sum_{j,k} \frac{\partial x^i}{\partial y_j} \frac{\partial x^k}{\partial y^j} V_k = G^{i,k} V_k$$

or in morphogenetic system where J is a rectangular matrix we have

$$V^k(x) = (J^T J)^{-1} V_k(x) = (J^T J)^{-1} J^T V(x) = G^{i,k} V_k.$$

The metric S is given by the expression

$$S^2 = V^k(x)^T V_k(x) = (G^{h,k} V_h(x))^T V_k(x) = V_h(x))^T (G^{h,k})^T V_k(x)$$

$$but \ (G^{h,k})^T = G^{h,k}$$

and

$$S^2 = V_h(x))^T (G^{h,k})^T V_k(x) = V_h(x))^T G^{h,k} V_k(x) = V_h(z)^T G^{h,k} V_k(z)$$

*and*

$$S^2 = V^h(x))^T G_{h,k} V^k(x) = V^h(z))^T G_{h,k} V^k(z)$$

When J is a rectangular matrix we can write the same expressions in this way

$$S^2 = ((J^T J)^{-1} J^T V(x))^T (J^T V(x)) = (J^T V(x))^T (J^T J)^{-1} (J^T V(x))$$

### 1.1 Commutators in morphogenetic system and tensor derivative

Given the operator D and the covariant components $J^T Y$ we have

$$DY_k = D(J^T Y) = J^T DY - J^T DY + D(J^T Y)$$
$$= J^T (DY) + (D(J^T Y) - J^T DY) = J^T (DY) + [D, J^T] Y$$

*where*

$$[D, J^T] = DJ^T - J^T D$$

*or*

$$J^T (DY) = D(J^T Y) - [D, J^T] Y$$

The covariant derivative of Y is equal to the derivative of the covariant components of Y minus the commutator of D with $J^T$ applied to Y. For the contro-variant components we have

$$DY^k = D((J^TJ)^{-1}J^TY) = (J^TJ)^{-1}J^TDY - (J^TJ)^{-1}J^TDY + D((J^TJ)^{-1}J^TY)$$

$$= (J^TJ)^{-1}J^T(DY) + (D((J^TJ)^{-1}J^TY) - (J^TJ)^{-1}J^TDY)$$

$$= (J^TJ)^{-1}J^T(DY) + [D,(J^TJ)^{-1}J^T]Y$$

*where*

$$[D,(J^TJ)^{-1}J^T] = D(J^TJ)^{-1}J^T - (J^TJ)^{-1}J^TD$$

*or*

$$(J^TJ)^{-1}J^T(DY) = DY^k - [D,(J^TJ)^{-1}J^T]Y$$

For example in the classical tensor calculus we have

$$[(J^TJ)^{-1}J^TD - D(J^TJ)^{-1}J^T]Y = (J^{-1}\frac{\partial}{\partial x_h} - \frac{\partial}{\partial x_h}J^{-1})Y_j = [J^{-1},\frac{\partial}{\partial x_h}]Y_j = \frac{\partial^2 x_k}{\partial x_h \partial y_j}$$

$$[J^TD - DJ^T]Y = (J^T\frac{\partial}{\partial x_h} - \frac{\partial}{\partial x_h}J^T)Y = [J^{-1},\frac{\partial}{\partial x_h}]Y = \frac{\partial^2 y_k}{\partial x_h \partial x_j}$$

These are the Christoffel elements for covariant and contravariant derivatives.

## 2. Quantum Mechanics from Vector of Boltzmann Entropies

In his paper *The statistic origin of quantum mechanics*, (Klein 2011) U. Klein takes into consideration 3 different levels of the role played by probability in Physics passing from the classical scenario to the quantum one:

*"With regard to the role of probability, three types of physical theories may be distinguished.*
*1. Theories of type 1 are deterministic. Single events are completely described by their known initial values and deterministic laws (differential equations). Classical mechanics is obviously such a theory. We include this type of theory, where probability does not play any role, in our classification scheme because it provides a basis for the following two types of theories.*
*2. Theories of type 2 have deterministic laws but the initial values are unknown. Therefore, no predictions on individual events are possible, despite the fact that deterministic laws describing individual events are valid. In order to verify a prediction of a type 2 theory a large number of identically prepared experiments must be performed. We have no problems to understand or to interpret such a theory because we know its just our lack of knowledge which causes the uncertainty. An example is given by classical statistical mechanics. Of course, in order to construct a type 2 theory one needs a type 1 theory providing the deterministic laws.*
*3. It is possible to go one step further in this direction increasing the relative importance of probability even more. We may not only work with unknown initial values but with unknown laws as well. In these type 3 theories there are no deterministic laws describing individual events, only probabilities*

*can be assigned. There is no need to mention initial values for particle trajectories any more (initial values for probabilistic dynamical variables are still required)."*

In the type 3 theories we have unknown laws so the entropy is dependent on the observer. In ordinary quantum mechanics we assume that the entropy is a vector that represents superposition state of the pure states. Any observer by a measure can see only one of the possible values of the superpose vector of entropies. Similarly to the work of L. del Rio et al. (Del Rio et al. 2011) we represent the different entropies as follows

$$S_j = S(A|B_j)$$

The $S_j$ is the conditional entropy of the system A conditioned to the observer $B_j$. Now the vector of the superpose entropies can be expressed in this way

$$\begin{cases} S_1 = -k \int \rho_1 \log \rho_1 dy \\ S_2 = -k \int \rho_2 \log \rho_2 dy \\ \quad \quad \ldots \\ S_n = -k \int \rho_n \log \rho_n dy \end{cases}$$

where

$$\rho_j = \rho_j(x_1, x_2, \ldots x_q, \theta_1, \theta_2, \ldots, \theta_p)$$

where x and θ are the variables and parameters of the probability distribution. The parameter k is the Boltzmann constant.

For the equilibrium condition the probability for any observer is equal in any position x and is function only of the parameters. So we have

$$\rho_j = \rho_j(x_1, x_2, \ldots x_q, \theta_1, \theta_2, \ldots, \theta_p) = \frac{1}{W_j(\theta_1, \theta_2, \ldots, \theta_p)}$$

And

$$\begin{cases} S_1 = k \log W_1(\theta_1, \theta_2, \ldots, \theta_p) \\ S_2 = k \log W_2(\theta_1, \theta_2, \ldots, \theta_p) \\ \quad \quad \ldots \\ S_n = k \log W_n(\theta_1, \theta_2, \ldots, \theta_p) \end{cases} \quad (2)$$

where W are the number of the microstates for the same parameters θ as temperatures, pressures, etc…. In this case S is a vector of Boltzmann entropies one for each observer. Now we have

$$J_{k,j} = \frac{\partial S_k}{\partial \theta_j} = \frac{\partial \log W_k}{\partial \theta_j}$$

and

$$F_{h,k} = J^T J = \sum_j \frac{\partial \log W_j}{\partial x_h} \frac{\partial \log W_j}{\partial x_k} \geq \Sigma^{-1}$$

where $F$ is the Fisher matrix (Braunstein & Caves, 1994). We have also that

$$F^{h,k} = (J^T J)^{-1} \leq \Sigma.$$

That is true for the Cramer Rao theorem for $\Sigma$ as a covariant matrix.

It is important to underline that the interpretation of quantum mechanics from a vector of Boltzmann entropies whose values are conditioned by the observer presents an important analogy with Rovelli's relational interpretation of quantum mechanics, indeed can be considered as a new elegant way to formulate relational quantum mechanics. As regards this view of quantum theory, the reader can find details in the references (Rovelli, 1996, 1997; van Fraassen, 2010; Bitbol, 2007). The central tenet of relational quantum mechanics is that there is no meaning in saying that a certain quantum event has happened or that a variable of the system $S$ has taken the value $q$: rather, there is meaning in saying that the event $q$ has happened or the variable has taken the value $q$ with respect to another system O. The apparent contradiction between the two statements that a variable has or has not a value is resolved by indexing the statements with the different systems with which the system in question interacts. Quantum events only happen in interactions between systems, and the fact that a quantum event has happened is only true with respect to the systems involved in the interaction. The unique account of the state of the world of the classical theory is thus fractured into a multiplicity of accounts, one for each possible "observing" physical system. In the words of Rovelli (Rovelli, 1996): "Quantum mechanics is a theory about the physical description of physical systems relative to other systems". On the basis of the mathematical formalism developed in this chapter, one can say that the superpose vector of the quantum entropies (whose values are conditioned by the observer) is the fundamental entity that explains in what sense, when a quantum event happens, it happens only with respect a peculiar observing physical system. In quantum mechanics, one can say that a variable of a quantum system takes the value $q$ because the vector given by a superposition of different Boltzmann entropies collapses to one specific entropy which corresponds to the value $q$ of the variable under consideration: an interaction between the quantum system and an observer happens which produces the collapse of the vector of the superposed entropies into one specific entropy and the observer involved in this interaction measures just the value $q$ of the variable under consideration.

## 3. From Vector of Boltzmann Entropies to Bohm's quantum potential

In chapter 2, we have shown that the morphogenetic calculus leads to a vector of superposition of different Boltzmann entropies that can lead to a new interesting reading of Rovelli's relational quantum mechanics. However, as we know, Rovelli's relational quantum mechanics seems an approach directed towards the reconciliation between irreversibility (collapse of the wave function and thus the standard interpretation of quantum mechanics) with the reversibility of the Lorentzian observers. Rovelli's approach indeed continuously turns each question on the collapse of the wave function. So, the wave function collapse remains the "fifth postulate" of quantum mechanics and diffuses the Von Neuman chain hidden dangers in a network of observers. In this regard, from the viewpoint of the standard interpretation of quantum mechanics and somewhat also of relational quantum mechanics, one can say that no collapse has happened until the information about the interaction between the quantum microsystem under consideration and the observer has arrived in this observer's brain. If an observer A has the information that the measurement has happened but an observer B has not had an interaction with the microsystem under consideration and thus has the information that the measurement has not happened, this implies that for the observer B the measured system and the brain of the first observer (A) have both remained in a superposition. According to Pavsic's view (Pavsic, 2001), this leads to the conclusion that "subjectively" a collapse of the wave function has occurred relative only to a peculiar observer's consciousness state, but "objectively" there is no collapse. Let us make the following example. Before an observer A measures the spin of an electron S, it is in a superposition state. Before this observer A has any contact with the electron, the apparatus O, or their environment, they are altogether in a superposition state. After looking at the apparatus, there is no longer superposition relative to his/her consciousness. However, relative to another observer B the combined state of the electron S, the apparatus O, and the brain of the observer A can remain in superposition until B himself gets in contact with A, O, S, or the environment of S, O, and A. As a consequence – Pavsic claims – it seems legitimate to argue that a wave function is always relative to something, or, better, to somebody. There can be no "objective" wave function. In the standard interpretation of quantum mechanics, and consequently in relational quantum mechanics too, a wave function can be always considered as relative to some observer. Thus the collapse of the wave function can be always considered as a subjective collapse. Relative to a given observer A, a wave function is collapsing all the time: whenever the information (direct or indirect—through the environmental degrees of freedom) about the outcome of measurement reaches the observer A. Instead, there is no collapse if we contemplate other observers B, C, D, …. performing their experiments. On the basis of these considerations, according to the authors of this article, it seems legitimate to conclude that – since an observer can always be defined for everything is in a superposition state – the quantum superposition and the quantum entanglement between given systems can be considered as the fundamental reality of the universe. But such aspect has not at all to be considered as "subjective": at the universe origin there were no observers; in a double slit experiment the electrons' pattern does not depend on an observer. What we mean to say is that the wave-function is not a mysterious ghost who vanishes when the measurement arrives, but something which acts on the matter and determines the conditions of the measurement itself (Cini & Levy Leblond, 1990). Quantum mechanics pictures the world also without observers: it is possible to eliminate the mystery of the collapse in a realistic picture by recognizing the contextual nature of the quantum measurement. In the second part of this paper, by making use of the morphogenetic calculus developed in chapter 1, we will provide a new interesting reading of Bohm version of quantum mechanics in which the quantum potential emerges as an information medium determined by the vector of the superposition of entropies (Licata, 2008; 2010).

In virtue of the formalism developed in chapter 1.1 and the transformation of the references

$$y_h = y_h(x_1, x_2, ..., x_n)$$

the covariant derivative of the vector Y in the coordinates x is

$$D^i_k Y = \frac{\partial Y_j}{\partial y^h} \frac{\partial x^i}{\partial y_j} \frac{\partial y^h}{\partial x^k} = \frac{\partial v^p}{\partial y^h} \frac{\partial y_j}{\partial x^p} \frac{\partial x^i}{\partial y_j} \frac{\partial y^h}{\partial x^k} + v^p \frac{\partial^2 y_j}{\partial y^h \partial x^p} \frac{\partial x^i}{\partial y_j} \frac{\partial y^h}{\partial x^k}$$

$$= \frac{\partial v^p}{\partial y^h} \frac{\partial y^h}{\partial x^k} \frac{\partial x^i}{\partial y_j} \frac{\partial y_j}{\partial x^p} + v^p \frac{\partial^2 y_j}{\partial y^h \partial x^p} \frac{\partial y^h}{\partial x^k} \frac{\partial x^i}{\partial y_j}$$

$$= \frac{\partial v^p}{\partial x^k} \frac{\partial x^i}{\partial y_j} \frac{\partial y_j}{\partial x^p} + v^p \frac{\partial^2 y_j}{\partial x^k \partial x^p} \frac{\partial x^i}{\partial y_j} = \frac{\partial v^p}{\partial x^k} + v^p \frac{\partial^2 y_j}{\partial x^k \partial x^p} \frac{\partial x^i}{\partial y_j}$$

Now for the transformation (1) we have the covariant derivatives

$$Dv = \frac{\partial v^p}{\partial x^k} + v^p \frac{\partial^2 S_j}{\partial x^k \partial x^p} \frac{\partial x^i}{\partial S_j} = (\frac{\partial}{\partial x^k} + \frac{\partial^2 S_j}{\partial x^k \partial x^p} \frac{\partial x^i}{\partial S_j}) v^p \quad (3)$$

where x are the average values of the position of the particle. The average is one of the parameters θ.

As regards equation (3) we have also

$$\frac{\partial^2 S_j}{\partial x^k \partial x^p} \frac{\partial x^i}{\partial S_j} = \frac{1}{\frac{\partial S_j}{\partial x^i}} \frac{\partial^2 S_j}{\partial x^k \partial x^p} = \frac{1}{\frac{\partial S_j}{\partial x^i} \frac{\partial S_j}{\partial x^h}} \frac{\partial^2 S_j}{\partial x^k \partial x^p} \frac{\partial S_j}{\partial x^h} = \frac{\frac{\partial^2 \log W_j}{\partial x^k \partial x^p}}{\frac{\partial \log W_j}{\partial x^i} \frac{\partial \log W_j}{\partial x^h}} \frac{\partial \log W_j}{\partial x^h}$$

When we have

$$\frac{\partial^2 \log W_j}{\partial x^k \partial x^p} = \frac{\partial \log W_j}{\partial x^i} \frac{\partial \log W_j}{\partial x^h}$$

we obtain

$$\frac{\partial^2 S_j}{\partial x^k \partial x^p} \frac{\partial x^i}{\partial S_j} = \frac{\frac{\partial^2 \log W_j}{\partial x^k \partial x^p}}{\frac{\partial \log W_j}{\partial x^i} \frac{\partial \log W_j}{\partial x^h}} \frac{\partial \log W_j}{\partial x^h} = \frac{\partial \log W_j}{\partial x^h} \quad (4)$$

and thus the deformation of the derivative for the non Euclidean geometry is given by the expression

$$\frac{\partial}{\partial x^k} + \frac{\partial^2 S_j}{\partial x^k \partial x^p} \frac{\partial x^i}{\partial S_j} = \frac{\partial}{\partial x^k} + \frac{\partial \log W_j}{\partial x_h} = \frac{\partial}{\partial x^k} + B_h \quad (5)$$

where $B_h$ is like Weyl gauge potential (Castro & Mahecha, 2005).

Now in the classical mechanics the equation of average motion can be expressed, by the definition of the action A, in this way

$$A = \int \rho [\frac{\partial A}{\partial t} + \frac{1}{2m} \frac{\partial A}{\partial x^i} \frac{\partial A}{\partial x^j} + V] dt d^n x \quad (6)$$

In quantum mechanics we have a deformation of the moments for the change of the geometry so we have:

$$A = \int \rho [\frac{\partial A}{\partial t} + \frac{1}{2m} (p_i + B_i)(p_j + B_j) + V] dt d^n x$$

$$A = \int \rho [\frac{\partial A}{\partial t} + \frac{1}{2m} (p_i p_j + B_i B_j) + V] dt d^n x$$

$$A = \int \rho [\frac{\partial A}{\partial t} + \frac{1}{2m} (p_i p_j + \frac{\partial \log W}{\partial x_i} \frac{\partial \log W}{\partial x_j}) + V] dt d^n x$$

$$= \int \rho [\frac{\partial A}{\partial t} + \frac{1}{2m} p_i p_j + V) dt d^n x + \frac{1}{2m} \frac{\partial \log W}{\partial x_i} \frac{\partial \log W}{\partial x_j})] dt d^n x$$

with the Euler Lagrange minimum condition we have that the Fisher information or quantum action assumes the minimum value when

$$\delta A = 0$$

*For*

$$\delta \int \rho [\frac{\partial A}{\partial t} + \frac{1}{2m} p_i p_j + V) dt d^n x + \delta \frac{1}{2m} \frac{\partial \log W}{\partial x_i} \frac{\partial \log W}{\partial x_j})] dt d^n x = 0$$

*so*

$$\frac{\partial A}{\partial t} + \frac{1}{2m} p_i p_j + V + \frac{1}{2m} (\frac{1}{W^2} \frac{\partial W}{\partial x_i} \frac{\partial W}{\partial x_j} - \frac{2}{W} \frac{\partial^2 W}{\partial x_i \partial x_j}) = \frac{\partial S}{\partial t} + \frac{1}{2m} p_i p_j + V + Q$$

(7)

where $Q$ is the Bohm quantum potential that is a consequence for the extreme condition of Fisher information.

On the basis of the latest equation (7), one can interpret the Bohm quantum potential as an information channel determined by the functions W defining the number of microstates of the physical system under consideration, which depend on the parameters of the distribution probability (and thus, for example, on the probability density of the wave function, namely, in a bohmian approach, the space-temporal distribution of an ensemble of particles, namely the density of particles in the element of volume $d^3x$ around a point $\vec{x}$ at time $t$) and which correspond to the vector of the superpose Boltzmann entropies. In other words, in the approach here suggested, the distribution probability of the wave function determines the functions W defining the number of microstates of the physical system under consideration, a quantum entropy emerges from these functions W given by equations (2), and these functions W, and therefore the quantum entropy itself (the vector of the superpose entropies), can be considered as the fundamental physical entities that determine the action of the quantum potential (in the extreme condition of the Fisher information) on the basis of equation

$$Q = \frac{1}{2m}\left(\frac{1}{W^2}\frac{\partial W}{\partial x_i}\frac{\partial W}{\partial x_j} - \frac{2}{W}\frac{\partial^2 W}{\partial x_i \partial x_j}\right) \quad (8).$$

In this way, inside the picture developed in this chapter, in the extreme condition of Fisher information, the Boltzmann entropies defined by equations (2) emerge as "informational lines" of the quantum potential. Bohm's quantum potential contains informational lines given by the functions W.

Now, in non-relativistic bohmian quantum mechanics, by introducing the definition (8) of the quantum potential inside the quantum Hamilton-Jacobi equation, we obtain

$$\frac{|\nabla S|^2}{2m} + V + \frac{1}{2m}\left(\frac{1}{W^2}\frac{\partial W}{\partial x_i}\frac{\partial W}{\partial x_j} - \frac{2}{W}\frac{\partial^2 W}{\partial x_i \partial x_j}\right) = -\frac{\partial S}{\partial t} \quad (9)$$

which provides an energy conservation law in quantum mechanics (Durr & Teufel, 2009).. In equation (9) two quantum corrector terms appear in the energy of the system, which are owed to the functions W linked with the quantum entropy. On the basis of equation (9), we can say that these two quantum corrector terms of the energy linked with the quantum entropy play an essential role in the quantum formalism because without them the total energy of the physical system would not be conserved. In other words, we can say that, in the approach here suggested, the distribution probability of the wave function determines the functions W defining the number of microstates of the physical system under consideration, a quantum entropy emerges from these functions W given by equations (2), and these functions W, (and thus also the quantum entropy given by equations (2)), in virtue of equation (9), produce two quantum corrector terms in the energy of the system. These two quantum corrector terms can thus be interpreted as a sort of modification in the background space, i.e. as a sort of degree of chaos of the background space determined by the ensemble of particles associated to the wave function under consideration. According to the approach here suggested, it is just the functions W and thus the quantum entropy (2) the fundamental entities which represent what are the geometric properties of space from which the quantum force and the behaviour of quantum particles, derive. The fact that, in non-relativistic bohmian mechanics, the quantum potential contains a global information on the environment in which the experiment is performed, and at the same time it is a dynamical entity - namely its information about the process and the environment is active - derives just from the functions W (and thus from the quantum entropy (2)) on the basis of equations (8) and (9). The functions W emerge just as informational lines of the quantum potential. It has to be mentioned here another advantage of Bohm Theory. What is usually defined as background space at quantum mechanics level is nothing but an "open door" on a typical level of quantum field theory; it is possible to show that the Feynman integral paths can be derived from Bohm quantum potential ( Vigier 1989; Fiscaletti 2011).

If we consider, for example, the double slit experiment, the quantum potential is an information potential, namely brings a global, instantaneous and active information on the process and its environment (in fact, it turns out to depend on the width of the slits, their distance apart and the momentum of the particle and turns out to modify the behaviour of the particle itself), because it is determined by the functions W, which represent its informational lines, and thus by the quantum entropy. So, if one of the two slits is closed the quantum potential changes just because its informational lines represented by the functions W change, and this information about the modification of these informational lines arrives instantaneously to the particle, which behaves as a consequence.

### 4. Some Remarks on Bohm Theory in Curved Space

The next important step is to see what happens to the quantum potential (8) in a relativistic version of Bohm theory in curved space-time. In this regard, according to the authors, it is interesting to take under consideration the bohmian approach to Klein-Gordon equation developed by F. Shojai and A. Shojai in the references (A. Shojai and F. Shojai, 2004) and (F. Shojai and A. Shojai, 2004), which has the merit to provide a geometrodynamic interpretation of relativistic Bohm theory characterized by a significant unification of the quantum and gravitational behaviours of matter in a geometric picture. In F. Shojai's and A. Shojai's model, the quantum Hamilton-Jacobi equation which derives from the decomposition of the wave function in its polar form $\psi = |\psi| \exp\left(\frac{iS}{\eta}\right)$ has the following form

$$\partial_\mu S \partial^\mu S = m^2 c^2 \exp Q \quad (10),$$

which is Poincarè invariant and has the correct non-relativistic limit and where the quantum potential is defined as

$$Q = \frac{\eta^2}{m^2 c^2} \frac{\left(\nabla^2 - \frac{1}{c^2}\frac{\partial^2}{\partial t^2}\right)|\psi|}{|\psi|} \quad (11).$$

According to the view suggested in this chapter, on the basis of equation (8), the space-temporal distribution of the ensemble of particles describing the individual physical system under consideration is assumed to generate a modification, and thus a degree of chaos, of the background space linked with the functions W and thus with the quantum entropy (2). This modification of the background space defined by the functions W determines a quantum potential which, taking into account F. Shojai's and A. Shojai's model, produces a quantum mass given by equation

$$M^2 = m^2 \exp Q \quad (12)$$

namely

$$M^2 = m^2 \exp \frac{1}{2m}\left(\frac{1}{W^2}\frac{\partial W}{\partial x_i}\frac{\partial W}{\partial x_j} - \frac{2}{W}\frac{\partial^2 W}{\partial x_i \partial x_j}\right) \quad (13).$$

Moreover, F. Shojai and A. Shojai have shown that, as regards Bohm's version of Klein-Gordon equation, by changing the ordinary differentiating $\partial_\mu$ with the covariant derivative $\nabla_\mu$ and by changing the Lorentz metric with the curved metric $g_{\mu\nu}$ inside equations (10), (11), it is possible to combine the Bohm quantum theory of motion and gravity and to interpret the quantum potential as the conformal degree of freedom of the space–time metric. In this picture, the quantum Hamilton-Jacobi equation of motion for a particle (of spin 0) in a curved background is the following:

$$g^{\mu\nu}\nabla_\mu S \nabla_\nu S = m^2 c^2 \exp Q \quad (14)$$

where

$$Q = \frac{\eta^2}{m^2 c^2} \frac{\left(\nabla^2 - \frac{1}{c^2}\frac{\partial^2}{\partial t^2}\right)_g |\psi|}{|\psi|} \quad (15)$$

is the quantum potential. By utilizing a fruitful observation of de Broglie (de Broglie, 1960), that the quantum theory of motion for relativistic spinless particles is very similar to the classical theory of motion in a conformally flat space-time in which the conformal factor is related to Bohm's quantum potential, the quantum Hamilton-Jacobi equation (14) can be written as

$$\frac{m^2}{M^2} g^{\mu\nu} \nabla_\mu S \nabla_\nu S = m^2 c^2 \quad (16).$$

From this relation (16) one can conclude that the quantum effects are equivalent to the change of the space-time metric from $g_{\mu\nu}$ to

$$\tilde{g}_{\mu\nu} = \frac{M^2}{m^2} g_{\mu\nu} \quad (17)$$

which is a conformal transformation. In this way equation (16) can be written as

$$\tilde{g}^{\mu\nu} \tilde{\nabla}_\mu S \tilde{\nabla}_\nu S = m^2 c^2 \quad (18)$$

where $\tilde{\nabla}_\mu$ represents the covariant differentiation with respect to the metric $\tilde{g}_{\mu\nu}$.

In this F. Shojai's and A. Shojai's model, the important conclusion we can draw is thus that the presence of the quantum potential is equivalent to a curved space-time with its metric being given by

$$\tilde{g}_{\mu\nu} = \frac{M^2}{m^2} g_{\mu\nu} \quad (17).$$

I. e., we have obtained that there is a geometrization of the quantum aspects of matter. It seems that there is a dual aspect to the role of geometry in physics. The space-time geometry sometimes looks like what we call gravity and sometimes looks like what we understand as quantum behaviours. In other words, in F. Shojai's and A. Shojai's model, the effects of gravity on geometry and the quantum effects on the geometry of space-time are highly coupled, the geometric properties which are expressed by the quantum potential and which determine the behaviour of a particle of spin zero are linked with the curved space-time: the particles determine the curvature of space-time and at the same time the space-time metric is linked with the quantum potential which influences the behaviour of the particles (Fiscaletti, 2012).

Now, in the approach suggested in this chapter based on the definition (8) of the quantum potential as physical entity characterized by information lines linked with the functions W namely with the quantum entropy, one can interpret in a clear way why and in what sense the quantum potential is the conformal degree of freedom of the space-time metric, why and in what sense the effects of gravity on geometry and the quantum effects on the geometry of space-time are highly coupled: the key of explanation of these results lies just in the functions W, and thus in the quantum entropy, corresponding with the degree of chaos, with the modification of the background space determined by the ensemble of particles associated with the wave function under consideration. The space-time which characterizes a relativistic curved space-time (in the presence of a particle of spin 0) has a conformal metric which is determined by the functions W (and thus is also linked with the quantum entropy) on the basis of equation

$$\tilde{g}_{\mu\nu} = g_{\mu\nu} \exp \frac{1}{2m}\left(\frac{1}{W^2}\frac{\partial W}{\partial x_i}\frac{\partial W}{\partial x_j} - \frac{2}{W}\frac{\partial^2 W}{\partial x_i \partial x_j}\right) \quad (19)$$

and the quantum Hamilton-Jacobi equation of motion (18) in this background becomes

$$g^{\mu\nu} \exp \frac{1}{2m}\left( \frac{1}{W^2}\frac{\partial W}{\partial x_i}\frac{\partial W}{\partial x_j} - \frac{2}{W}\frac{\partial^2 W}{\partial x_i \partial x_j}\right) \tilde{\nabla}_\mu S \tilde{\nabla}_\nu S = m^2 c^2 \quad (20).$$

The dual role of the geometry in physics expected by F. Shojai's and A. Shojai's model receives therefore a new suggestive reading: the real key of reading of the link between gravity and quantum behaviours lies just in the functions W, namely in the quantum entropy (2). The effects of gravity on geometry and the quantum effects on the geometry of space-time are highly coupled because they are both determined by the background space described by the quantum entropy, they are both produced by the modification of the background space supporting the density of particles associated with the wave function under consideration which is contained in the functions W.

On the basis of the picture proposed in this chapter, one can say that the quantum particles determine the curvature of space-time and at the same time the space-time metric is linked with the functions W indicating the informational lines of the motion of the particles. The functions W, the informational lines of the quantum potential, and thus the quantum entropy, appear indeed as real intermediaries between gravitational and quantum effects of matter. The functions W, and thus the quantum entropy generates itself a curvature which may have a large influence on the classical contribution to the curvature of the space-time.

## 5. Conclusions

In this paper we have provided a new suggestive reading of quantum mechanics by starting from the morphogenetic calculus and from the vector of superposition of different Boltzmann entropies. In the first part of the article, we have seen that, while in classical thermodynamics entropy is defined independently from the observer, in quantum mechanics a system can be defined by its entropy whose value is conditioned by an observer. The collapse of the vector given by a superposition of different quantum entropies into a specific entropy as a consequence of an interaction with an observer introduces a new suggestive reading of Rovelli's relational interpretation of quantum mechanics.

In the second part of the article, given the relation between the number of the microstates W function of the macrostates and given the entropies S as Cartesian coordinates, the reference can be changed and from Euclidean coordinates of the entropies, in which the observers are independent, one can move to a non Euclidean space of the parameters (averages, variances …. ) . In this background we compute covariant derivatives in the parameter space and we can obtain by entropy the Bohm quantum potential and the quantum effects. The approach suggested in this background leads to the idea that, in the extreme condition of Fisher information, Bohm's quantum potential emerges as an information channel determined by informational lines associated with the vector of the superpose Bolzmann entropies. This approach has the merit that, in a relativistic curved space-time, the informational lines associated with the quantum entropy appear as real intermediaries between gravitational and quantum effects of matter, determining a high coupling between the effects of gravity on geometry and the quantum effects on the geometry of space-time.